\UseRawInputEncoding
\documentclass[aip,graphicx]{revtex4-1}
\usepackage{xcolor}
\usepackage{graphicx}

\usepackage[english]{babel}

\usepackage{comment}
\usepackage[dvipsnames]{xcolor}
\usepackage{empheq}
\usepackage{amsmath}
\usepackage{bbold}
\usepackage{amssymb}
\usepackage{ulem,xpatch}
\usepackage{tabularx}
\usepackage{physics}
\usepackage{siunitx}
\usepackage{soul}

\usepackage{lineno}

\newcommand{\degree}{\ensuremath{^\circ}}

\newcommand{\hide}[1]{\relax}

\newcommand{\nocontentsline}[3]{}
\newcommand{\tocless}[2]{\bgroup\let\addcontentsline=\nocontentsline#1{#2}\egroup}

\draft 

\begin{document}


\title{Wide-field stroboscopic imaging of topologically protected phononic modes} 



\author{
    Ilia Chernobrovkin$^{1,2,*}$,
    Maurice Debray$^{1,2,*}$,
    Frederik Holst Knudsen$^{1,2,*}$,
    Thibault Capelle$^{1,2}$,
    Mads Bjerregaard Kristensen$^{1,2}$, 
    Michael Pitts$^{1,2}$,
    Xiang Xi$^{1,2,\dagger}$,
    Albert Schliesser$^{1,2,\dagger}$}

\address{$^{1}$\textnormal{Niels Bohr Institute, University of Copenhagen, 2100 Copenhagen, Denmark}\\
$^{2}$\textnormal{Center for Hybrid Quantum Networks (Hy-Q), Niels Bohr Institute},\\ \textnormal{University of Copenhagen, 2100 Copenhagen, Denmark}\\
{$^\star$}\textnormal{These authors contributed equally}\\
{$^\dagger$}\textnormal{Corresponding authors}
}
\email{{$^\dagger$}\textnormal{xiang.xi@nbi.ku.dk, albert.schliesser@nbi.ku.dk}}


\date{\today}

\begin{abstract}

Imaging spatial mode profiles is important for understanding the behavior of mechanical resonators. The recent development of phononic circuits has increased the demand for a fast imaging method based on principles of coherent detection. However, it becomes complicated to perform measurements on a large surface area.
Here, we present a frequency-detuned collimated-beam interferometry measurement scheme with in-plane spatial resolution of about $\SI{6}{\micro\meter}$, which can provide information about the phase dynamics of the entire mechanical oscillation cycle on a time scale of a few seconds. 
We employ a stroboscopic pulse probing method to resolve high-frequency vibrational motion with a standard CMOS camera.
We use this setup to image megahertz frequency resonant mode profiles present in a Valley-Hall topological triangular cavity, over an area of more than $20\, \mathrm{mm}^2$. 
%
%
We relate the obtained data to numerical simulations of the topological edge modes to reveal the relation between backscattering and the mode profile distribution. 
The presented protocol can become a staple for characterizing mesoscopic mechanical resonators.
\end{abstract}


\maketitle 

\section{Introduction}
Nano- and micromechanical devices have been extensively studied for various applications, including timing, sensing, microwave signal processing, and quantum optomechanics \cite{RMPCOM2014,RMPmechanics2022}.
Visualization of mechanical mode profiles offers direct experimental evidence of elastic energy distribution and can be used, in combination with other techniques, to infer some properties of the system. 
It can reveal the spatial confinement of the mode, its coupling to other oscillators, and geometrical features such as symmetry and chirality\cite{ma2021nn, ma2021am}. 
Traditionally, two main experimental protocols have been used for obtaining spatial profiles of mechanical motion. One is based on optical interferometry with a focused laser beam\cite{Cha2018nature, ma2021nn,yan2018nm, barg_measuring_2017, romero_propagation_2019}, another is transmission-mode microwave impedance microscopy on an atomic force microscopy platform\cite{zhang2022ne}. 
To obtain the imaging data, both of them rely on point-to-point raster scan over the device surface, which can take several hours even with a rough spatial resolution. 
Long acquisition time introduces errors related to drift present in the system, which makes matching different data points and interpreting the resulting image more complicated.
A faster dark-field method typically suffers from poor sensitivity\cite{barg_measuring_2017}.

Here, we demonstrate a fast imaging technique for obtaining profiles of mechanical waveguide modes over a large area.
This method is based on the principles of stroboscopic pulsed light probing, which allows to detect high-frequency periodic motion using a standard CMOS camera\cite{Hart2000}. This concept has already been implemented for various applications and research in microelectromechanical (MEMS) and biological systems\cite{Conway2017,Shavrin2013oe,Heikkinen2013oe,kokkonen2015oe,Iimori2022oe}. 
We adapt it for use in an optomechanical interferometer setup and introduce a frequency-detuned pulse train scheme for detection of dynamic motion of devices with one measurement cycle in only 2 seconds. The detuned-frequency scheme enables us to differentiate the mechanical phase of a chosen mode across a large surface area of the device. As a result, we acquire the flexural motion magnitude and phase over the area of 20 mm$^2$. 

We apply the stroboscopic imaging technique to topological valley-Hall  phononic waveguide systems. The topological systems are presented in our previous work\cite{xi2024soft}, whose concept originates from condensed matter physics and, over the past decades, has been extended into the field of nanomechanics \cite{Lu2017np,Cha2018nature,yan2018nm,yu2018nc,ma2021am,ma2021nn, xi2021sa, ren2022nc, zhang2022ne, Shah2024col}. 
Topological waveguides emerge at the interface of two bulk crystals with distinct topological phases, and feature wave transport around sharp bends with very low suppressed backscattering\cite{xi2024soft}, which is desired for complex integrated phononic and photonic circuits.
In experimental studies, the key research objectives include confirming the localization of topological edge states, quantification of their losses and their topological protection \cite{Cha2018nature,Parappurath2020sa,xi2024soft}. Imaging the displacement field of the modes offers direct experimental evidence for addressing some of these research objectives.
Besides, the images we acquire through the stroboscopic probing technique for the first time provide visual evidence that large (small) backscattering leads to modified (normal) mode profiles.
\section{Experiments}

\begin{figure}
\centering
\begin{center}
\includegraphics[width=0.7\linewidth]{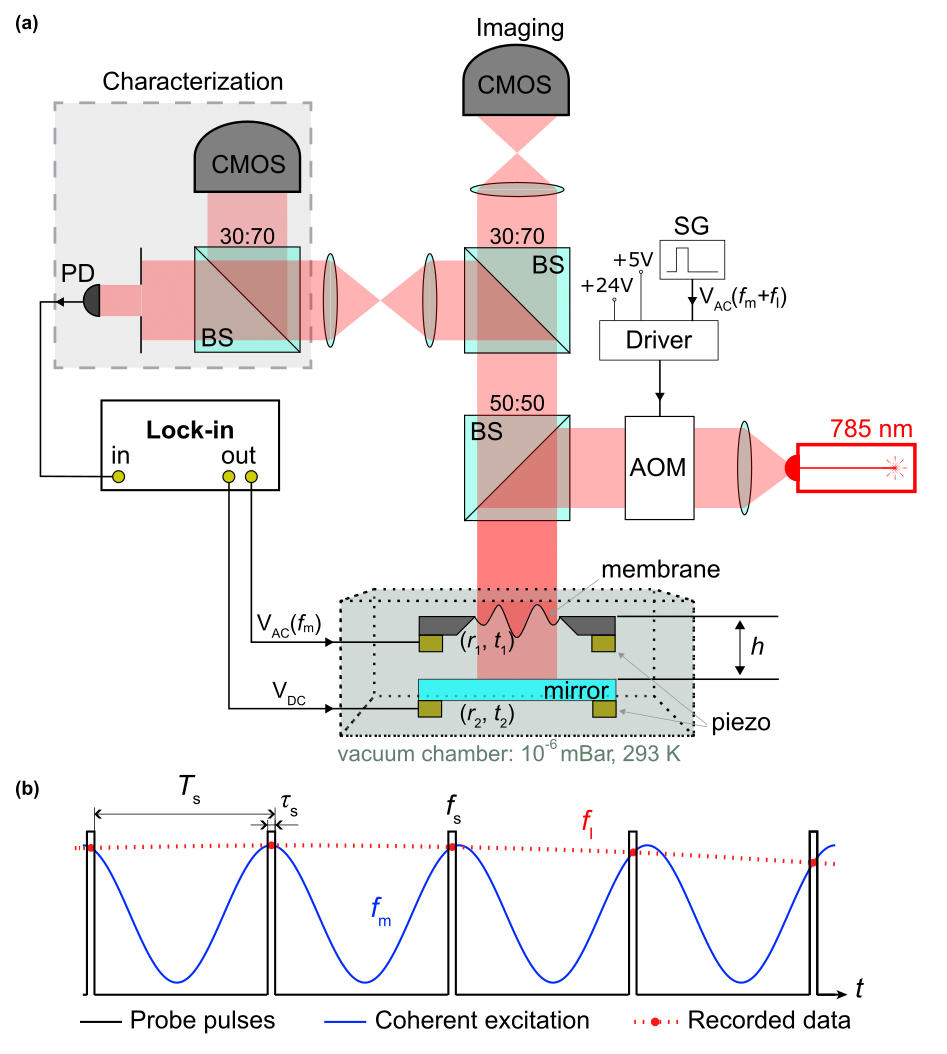}
\caption{ \textbf{Experimental protocol.}
(a) Experimental setup for stroboscopic imaging measurements. Depicted here, AOM: acousto-optic modulator, BS: optical beam splitter with noted R:T values, CMOS: complementary metal–oxide–semiconductor camera, Lock-in: lock-in amplifier, PD: silicon photodetector, Piezo: piezoelectric actuator, SG: signal generator. The dashed characterization box is the setup for measuring the mechanical spectra of a targeted region. 
(b) Detuned stroboscopic detection scheme: the detecting optical pulses (black line) has a repetition frequency ($f_\textrm{s}$) slightly different from the mechanical excitation (blue line) frequency ($f_\textrm{m}$). 
\label{f:fig2_setup} }
\end{center}
\end{figure}

Our measurement protocol combines the principles of the stroboscopic effect, optical interferometry, and wide-field camera imaging. This resulted in the setup configuration presented in Fig. \ref{f:fig2_setup}a. 
%
%
%
At the heart of the system we have a Fabry–Pérot (FP) interferometer formed by an vibrating device membrane and a microscope mirror, which are separated by a vacuum gap $h$. The gap $h$ can be tuned by a piezoelectric transducer positioned below the mirror.
The membrane-mirror system is placed inside a vacuum chamber kept at $10^{-6}$ mBar at room temperature.
We create stroboscopic pulses by threading a collimated laser beam (wavelength $\lambda$ = 785 nm) through an acousto-optic modulator. The pulse train frequency and duty cycle are defined by an external signal generator. The interfered light that escapes from the FP interferometer is sent to a CMOS camera that forms the image, while a small portion of the beam is directed into a "characterization" arm of the interferometer (\ref{f:fig2_setup}a, gray-shaded area). The characterization arm consists of a photodetector with a pinhole placed in front of it and serves to detect resonant motion at a specific spot of the membrane surface.

%

We consider the optical intensity measured by the imaging camera $I_\textrm{out}(x,y,t)$, where $(x,y)$ is a spatial coordinate of the vibrating membrane surface.
We denote the membrane and mirror reflection (transmission) amplitudes as $r_1$ ($t_1$) and $r_2$ ($t_2$) respectively. The coefficients $r_1$ and $t_1$ are dependent on $(x, y)$, while $r_2$ and $t_2$ are taken to be uniform across the imaging region and therefore spatially independent. For simplicity we consider the amplitudes to be real.
Due to the small value of the reflection $r_1$ and $r_2$, we ignore the internal resonance enhancement effect from the FP cavity. 
Given the light intensity $I_\textrm{in}(t)$ injected into the chamber, the light intensity $I_{out}$ detected at the imaging camera would therefore be:
\begin{equation} \label{eq:output}
    \begin{split}
    I_\textrm{out}(x,y,t) = \beta I_\textrm{in}(t)\cdot[(t_1^2 r_2e^{i\cdot2k\cdot[h+u(x,y,t)]}+r_1)\cdot (t_1^2 r_2e^{i\cdot2k\cdot[h+u(x,y,t)]}+r_1)^*] \\
    =\beta I_\textrm{in}(t)\cdot[ \left|t_1^2 r_2\right|^2+\left|r_1\right|^2+2\left|t_1^2 r_2 r_1\right| \cos{(2k[h+u(x,y,t)])}],
    \end{split}
\end{equation}
Here $k=2\pi/\lambda$ is the laser beam wavenumber, $\beta$ is the optical attenuation in the path, 
$u(x,y,t)=u_\textrm{A}u_z(x, y) \allowbreak \cdot \phi(t) $ is the out-of-plane component of the mechanical vibration with the amplitude $u_\textrm{A}\ll\lambda$. $u_z(x, y)$ is the normalized mode profile. For a standing wave mechanical mode $\phi(t)=\cos{(2\pi f_\textrm{m} t + \varphi) }$, $f_\textrm{m} \sim1.25$ MHz is the mechanical actuation frequency, $\varphi$ is a tuneable phase controlled by excitation. 
In the experiment, to enhance the detection efficiency, the gap $h$ is tuned by a piezo-actuator to satisfy the condition $\cos(2kh) \approx 0$. In this case, the output in Eq.\eqref{eq:output} can be simplified as: 
%
\begin{equation} \label{eq:output_simp}
    I_\textrm{out}(x,y,t) \approx \beta I_\textrm{in}(t)\cdot[\left|(t_1^2 r_2)\right|^2+\left|r_1\right|^2+\left|t_1^2 r_2 r_1\right|\cdot2k\cdot u(x,y,t)].
\end{equation}
In our experiment, the total peak power of input optical pulses onto the membrane is $\sim \SI{600}{\micro\watt}$. 

The input intensity $I_\textrm{in}$ has a repetition frequency of $f_\textrm{s} = 1/T_\textrm{s}$ and a duty cycle of $\tau_\textrm{s}/T_\textrm{s} = 5\%$. 
%
%
As such, the probe optical pulses $I_{in}(t)$ can be expanded through a Taylor series as $I_\textrm{in}(t) = \sum_{n=0}^{+\infty}I_n^\mathrm{in} \cos (2\pi n f_s t)$. 
%
%
By applying this definition to Eq. \ref{eq:output_simp}, we obtain an expression that contains products between the pulse harmonics and the mechanical response $u(x,y,t)$. Due to slow response of the CMOS imaging sensor, all high-frequency terms are averaged out, and we are left with one slow-varying component on top of a constant background:
\begin{equation} \label{eq:mechanical}
    I_\textrm{m}(x,y,t) = \beta I_1^\textrm{in}k\left|t_1^2 r_2 r_1\right|\cdot  u_\mathrm{A}u_z(x, y)\cos{[2\pi f_\textrm{I} t + \varphi]} + \beta I^\textrm{in}_0\cdot(\left|(t_1^2 r_2)\right|^2+\left|r_1\right|^2),
\end{equation}
which oscillates at the frequency $f_\textrm{I} = f_\textrm{s} - f_\textrm{m}$.
In most other camera-based experiments, the stroboscopic pulse sequence and the mechanical resonant excitation have the same frequency ($f_\textrm{I} = 0$), and the intensity $I_\textrm{m}$ is independent of time \cite{Hart2000}. In this case, different phases of the dynamic motion of $u_z(x, y)$ are obtained by repeating the image acquisition process at different phase delays ($\varphi$), which requires a stable phase lock between the pulses and the mechanical excitation.
%
%
We instead design a single-measurement-cycle protocol to reconstruct the full mechanical motion waveform by introducing a slight frequency offset $f_\textrm{I} = f_\textrm{s} - f_\textrm{m} \ne 0$, between the pulse repetition and the mechanical excitation frequencies, as shown in Fig. \ref{f:fig2_setup}. The dynamic motion can be extracted from a single time-resolved acquisition procedure. This constitutes the main defining feature of the stroboscopic imaging protocol that we employ.
In our experiment, $f_\textrm{I} = 0.5$ Hz, which is sufficient for imaging with a camera operating at the shutter frequency $f_\textrm{c} = 20$ Hz.
%
 %
Taking an arbitrary time $t_0$ as a reference, the $N$'th camera frame $P(x,y, N)$ contains the result of displacement integration over 1 of 40 (=$f_\textrm{I}/f_\textrm{c}$) mechanical periods: 
\begin{equation} \label{eq:frames}
   P(x, y, N) \propto \int_{t_0+(N-1)/f_\mathrm{c}}^{t_0+N/f_\mathrm{c}} I_{\mathrm{m}}(x, y, t)\, dt.
\end{equation}
Therefore, 40 consecutive camera image frames can capture dynamic mechanical displacement over one complete oscillation cycle. 
Taking the dynamic motion $ P(x_{\mathrm{ref}},y_{\mathrm{ref}},N) (N=1\rightarrow40)$ at one representative pixel as a reference waveform, and comparing it with the waveform $P(x_{i},y_{i}, N)$ at the other pixels, one can extract the relative mechanical phases at pixels over the entire image area. 
%
%
Practically, we perform several data acquisition cycles at different membrane-to-mirror gaps $h$ to ensure that we get the best signal-to-noise ratio, where $\cos(2kh) \approx 0$, for each imaged spot (for a non-zero tilt between the membrane and the mirror, $h$ itself depends on $x$ and $y$). 
The total imaging time is around three minutes plus around ten minutes of data transfer time from the camera to the computer. 

\begin{figure}
\centering
\begin{center}
\includegraphics[width=0.95\linewidth]{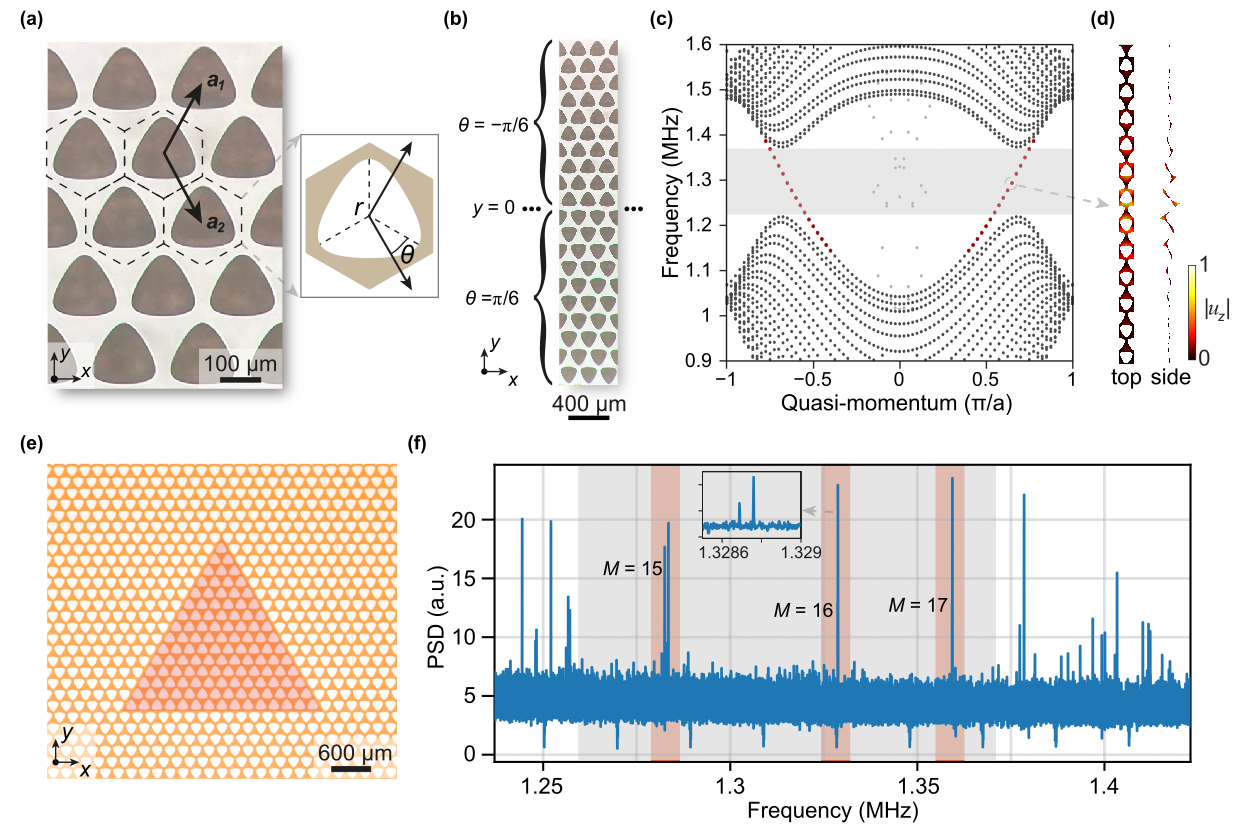}
\caption{\textbf{Topological phononic waveguide.}
(a) Microscope image of fabricated bulk valley-Hall topological crystals.
Dashed lines mark out unit cell of the honeycomb lattice with the basis vectors $\mathbf{a}_1$ and $\mathbf{a}_2$.
The right panel shows a single unit cell with a rounded triangular hole, whose orientation is defined by the angle $\theta$ and a size is defined by $r \approx \SI{93}{\micro\meter}.$
(b) Microscope image of the topological edge by interfacing two bulk insulators with $\theta = +\pi/6$ and $\theta = -\pi/6$. 
(c) Simulated band diagram of the edge structure in (b). The black, red, and grey dots respectively represent topological bulk, topological edge, and trivial in-plane mechanical modes.
(d) Simulated mode profile of the topological edge state. $u_z$ is the out-of-plane mechanical field component.
(e) Microscope image (false color) of fabricated topological triangular cavity.
(f) Meaured normalized power spectral density (PSD) from one of the fabricated devices. Inset shows the zoomed-in spectrum for the mode pair with $M = 16$.  } 
\label{f:fig1} 
\end{center}
\end{figure}

In this work, we apply the imaging protocol to topological mechanical valley-Hall insulators, which follow our previous design based on thin silicon nitride (SiN) membranes\cite{xi2024soft}, as shown in Fig. \ref{f:fig1}. The membrane has a thickness of 50 nm and a tensile stress $\sim1.27~\mathrm{GPa}$. 
The mechanical flexural modes vibrating out of the membrane plane are of interest.
The valley-Hall insulators are imitated by periodically patterning the membranes with rounded triangular holes on a honeycomb lattice with basis vectors $\mathbf{a}_1$ and $\mathbf{a}_2$ ($|a_1|=|a_2|= \SI{200}{\micro\meter}$) (Fig. \ref{f:fig1}a). 
Inside each unit cell, the angle $\theta$ between the triangle's symmetry axis and a lattice vector (Fig. \ref{f:fig1}a) defines the symmetries of the geometry (see Ref. \cite{xi2024soft} for details). 
%
This crystal configuration exhibits properties of the valley-Hall effect and features a non-trivial valley-Hall phase\cite{Lu2014_prb_Dirac_cones,lu2015prl}.
Reversing the incision orientation angle from $\theta$ to $-\theta$ preserves the energy band structure while inverting the valley-Hall phase.

We obtain localized topological mechanical waveguides by combining two bulk crystals with $\theta = \pi/6$ and $\theta = -\pi/6$ with a zigzag interface (Fig.~\ref{f:fig1}b). 
Due to the bulk-boundary correspondence, there exists a localized topological chiral edge mode within the bandgap at each valley, which is confirmed by our numerically simulated band diagram and mode profile shown in Figs.~\ref{f:fig1}c,d, respectively.
We wrap the topological waveguide into a closed-loop triangular cavity, as shown in the device picture in Fig. \ref{f:fig1}e. 
This configuration supports whispering gallery modes propagating along a closed path near the domain interface. 
Especially, the modes with longitudinal wave numbers $k^\parallel_M=2\pi M/L$ will be in-resonance due to the periodic boundary condition of the closed path, where $M$ is integer and $L\approx 9{.}3~\mathrm{mm}$ is the round-trip length of the waveguide. 
We measure the power spectral density at a point near the domain interface from one of the fabricated devices, as shown in Fig. \ref{f:fig1}f. 
We observe a phononic bandgap (gray-shaded region) with three pairs of resonant modes, whose mode numbers $M$ are 15, 16, 17, respectively.
The frequency splitting of the two modes for each pair is the result of backscattering induced coupling between clockwise and counter clockwise propagating topological edge modes, and can be utilized to characterize the strength of backscattering \cite{ren2022nc,xi2024soft,Shah2024col}.
%

%

\begin{figure}
\centering
\begin{center}
\includegraphics[width=\linewidth]{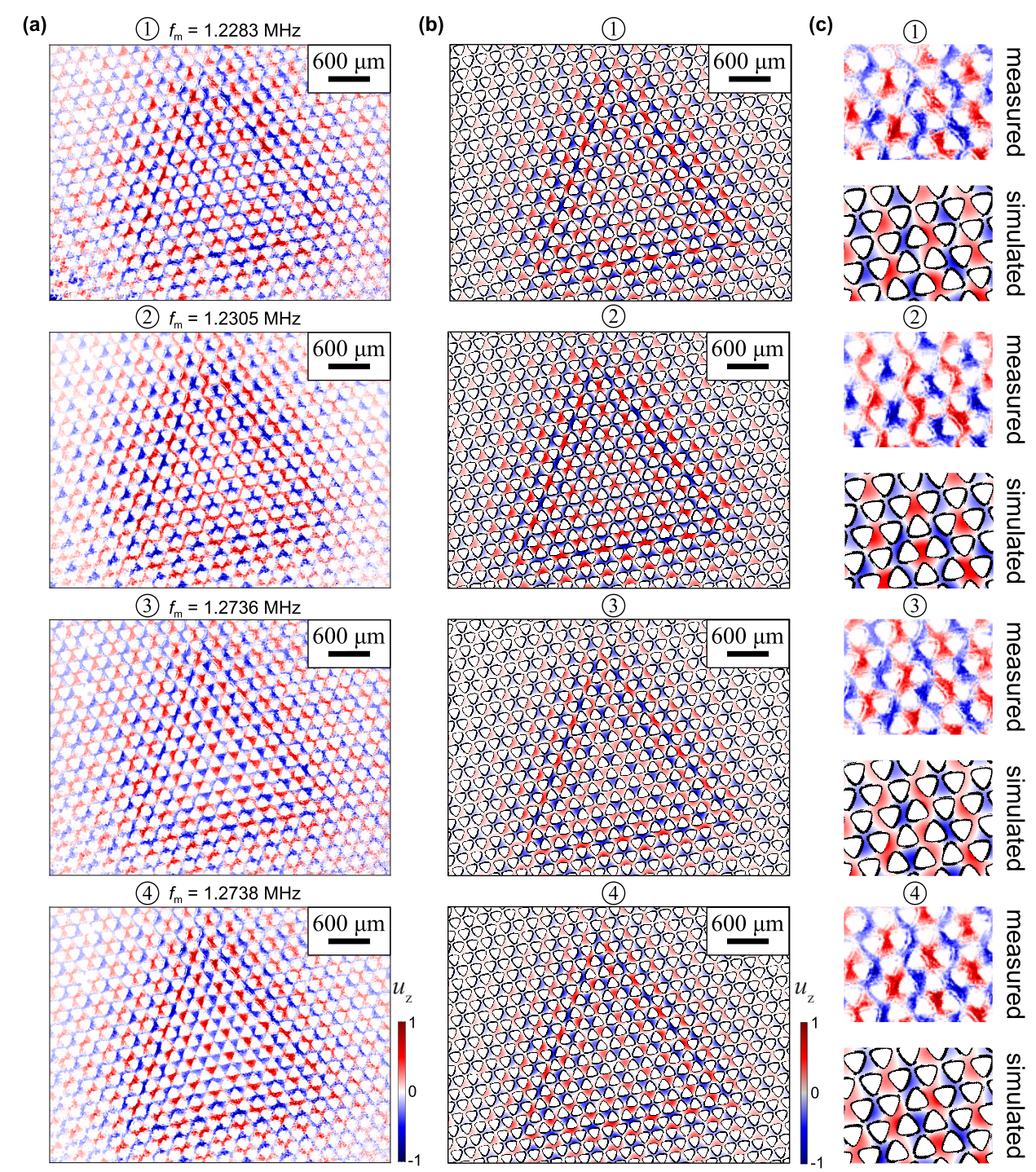}
\caption{\textbf{Resonant mode profiles.} 
(a) Experimental mode profiles of $u_z$ from stroboscopic method for the two topological mode pairs 1, 2 and 3, 4 from one device. Mode 1 and 2 have mode number of 15 while mode 3 and 4 have mode number of 16.
(b) Simulated modal profile for the mode 1, 2, 3, 4.
(c) Zoomed-in comparison for the measured and simulated results around the top triangle corner in (a) and (b). }
\label{f:fig3_results} 
\end{center}
\end{figure}

We image the displacement fields of the topological modes, as presented in Fig. \ref{f:fig3_results}a. 
For this demonstration we select two pairs of split modes from one of the devices. 
All obtained images (Fig. \ref{f:fig3_results}a) show the resonant topological edge modes confined around the triangular edge with very clear modal profile details.
The measured dynamic mechanical displacement over one complete oscillation cycle allows us to extract the relative mechanical phase of all the spatial pixels.
In the displacement maps of Fig. \ref{f:fig3_results}a, the mechanical vibrations with phase 0 $(u_z(x, y)>0)$ are colored in red, while the regions with phase $\pi$ $(u_z(x, y)<0)$ are depicted in blue. 
The spatial resolution of the images is $\sim \SI{6}{\micro\meter}$, limited by the numerical aperture of the focusing lens and the wavelength of the light source. 

The modes noted as 1 and 2 in Fig. \ref{f:fig3_results}a, corresponding to the coupled clockwise and counter-clockwise waves with the angular mode number $M = 15$, have a large frequency splitting of $\Delta_{15}\sim 2.2$ kHz. This coupling is due to the backscattering in the closed-loop waveguide, whose strength is characterized by the frequency splitting \cite{xi2024soft}. The modes pair 3 and 4 with $M = 16$ are separated by a small frequency of 200 Hz, which aligns well with observations in analogous device designs\cite{xi2024soft}. 
%

We perform numerical simulations of the imaged device with commercial finite element method. From the results we extract mode profiles of the pairs that, like the ones obtained experimentally, pertain to the edge eigenmodes with $M=15$ and $M=16$, respectively. We present them in Fig. \ref{f:fig3_results}b side-by-side with the corresponding experimental images for comparison.
All obtained profiles exhibit the similar main features as their simulated counterparts, showcasing an imaging resolution sufficient to resolve sub-wavelength details. Modes 3 and 4 exhibit almost perfect correspondence in the mode displacement distribution between the experimental results and the numerically predicted profiles.
%
%
%
%
In contrast to that, the measured profiles of modes 1 and 2 are different compared to their simulated counterparts, especially the distribution of the standing wave nodes and antinodes along the triangular interface. 
To highlight this, we show sections of the mode profiles near one of the corners of the triangle waveguide (Fig. \ref{f:fig3_results}c). 
We observe that for modes 1 and 2 neither a node nor an antinode appears aligned with the triangular corner vertex in the experimentally obtained images. This detail does not agree with simulations and was not observed in devices that we have presented in the previous work\cite{xi2024soft}. 
That is an indication of interference induced by contamination or fabrication defects near the waveguide, which is dominant over that from corner scattering. Such an abnormal mode pattern correlates with the strong scattering suggested by a large frequency splitting.
%
%
%

For comparison, mode profiles 3 and 4 show good agreement on displacement distribution between the experimental results and the numerically
predicted profiles, respectively.
They also have relatively low frequency splitting, indicating correspondingly lower backscattering.
The evidence is also supported by the fact that the pairs of modes presented in our previous work also show significantly lower splitting values\cite{xi2024soft}.
%
%


\section{Conclusion}
The combination of stroboscopic pulse probing, collimated-beam interferometry and wide-field camera imaging constitutes a strong tool for characterization of mechanical displacement in mesoscopic membrane resonators. We demonstrated its potential by applying the detuned stroboscopic imaging protocol to a valley-Hall topological waveguide and obtained profiles of two pairs of standing edge modes and for the first time observed the correlation between modal profile distortion and frequency splitting. The optical diffraction-limited resolution of $\sim \SI{6}{\micro\meter}$ allowed us to resolve subwavelength features of mechanical mode profiles, while the interferometric configuration made it possible to extract the flexural displacement phase. The CMOS camera-based detection scheme shows a significant improvement in terms of data acquisition time compared to canonically used raster scan techniques. The experimental results can be directly compared to numerical simulations and offer additional information about the properties of mechanical resonators. The setup scheme, data acquisition and analysis protocols can be readily integrated into larger, systematic optomechanical characterization protocols, and help observe interesting phenomena in adjacent fields, like topological phononics.

\paragraph*{\textup{\textbf{Acknowledgements}}}
This work was supported by the European Research Council project PHOQS (Grant No. 101002179), the Novo Nordisk Foundation (Grant Nos. NNF20OC0061866 and NNF22OC0077964), the Danish National Research Foundation (Center of Excellence “Hy-Q”), European Union's Horizon 2020 research and innovation programme under the Marie Skłodowska-Curie grant agreement No. 101107341, and a research grant (VIL59143) from the Villum Foundation.

\paragraph*{\textup{\textbf{Competing interests}}}
A. S. is a co-founder of the company Qfactory ApS, which commercializes soft-clamped phononic resonators. The other authors declare no competing interests.

\paragraph*{\textup{\textbf{Author contributions}}}
M.D. and F.H.K devised the experimental setup, data acquisition and analysis protocols with contributions from I.C., M.B.K. and T.C.
I.C. obtained and analyzed the experimental data.
T.C. and M.P. fabricated the device.
X.X. and M.B.K. conducted numerical simulations.
X.X. created the device design.
I.C., X.X. and Albert S. wrote the manuscript with contribution from all the authors.
X.X. and Albert S. supervised the project.

\bibliography{strobe_reference}

\end{document}